\documentclass{aa}
\usepackage{natbib}
\usepackage{epsfig}
\usepackage{csquotes}
\usepackage{tikz}
\usetikzlibrary{decorations.pathmorphing}
\usetikzlibrary{decorations.pathmorphing,calc}
\def\ut#1{\mathop{\vtop{\ialign{##\crcr
     $\hfil\displaystyle{#1}\hfil$\crcr\noalign
     {\kern1pt\nointerlineskip}\hbox{$\hfil\sim\hfil$}\crcr
     \noalign{\kern1pt}}}}}

\def\undersymbol#1#2{\mathop{\vtop{\ialign{##\crcr
     $\hfil\displaystyle{#2}\hfil$\crcr\noalign
     {\kern1pt\nointerlineskip}\hbox{$\hfil#1\hfil$}\crcr
     \noalign{\kern1pt}}}}}

\usepackage{graphicx}
\begin{document}

\title{Cosmic voids and the kinetic analysis. \\ II. Link to Hubble tension}
       \author{V.G.Gurzadyan\inst{1,2}, N.N.Fimin\inst{3}, V.M.Chechetkin\inst{3} }

              \institute{Center for Cosmology and Astrophysics, Alikhanian National
Laboratory and Yerevan State University, Yerevan, Armenia \and
SIA, Sapienza Universita di Roma, Rome, Italy \and Keldysh Institute of Applied Mathematics of RAS, Miusskaya Sq., 4, Moscow, Russia}

   \offprints{V.G. Gurzadyan, \email{gurzadyan@yerphi.am}}
   \date{Submitted: XXX; Accepted: XXX}

 \abstract{We consider a principal problem, that of the possible dominating role of self-consistent gravitational interaction in the formation of cosmic structures: voids and their walls in the local Universe. It is in the context of the Hubble tension as a possible indication of the difference in the descriptions of the late (local) and early (global) Universe. The kinetic Vlasov treatment enables us to consider the evolution of gravitating structures where the fundamental role has the modified gravitational potential with a cosmological constant, leading to the prediction of a local flow with a Hubble parameter that is nonidentical to that of the global Hubble flow. The Poisson equation for a potential with an additional repulsive term, including an integral equation formulation, is analyzed, and we predict the appearance of multiply connected two-dimensional gravitating structures and voids in the local Universe. The obvious consequence of the developed mechanism is that the cosmological constant poses a natural scaling for the voids, along with the physical parameters of their local environment, which can be traced in observational surveys.
}


   \authorrunning{N.N.Fimin, V.M.Chechetkin, V.G. Gurzadyan}
   \titlerunning{Filaments and the kinetic analysis}
   \maketitle
%
%________________________________________________________________

\section{Introduction}

The recently emerged Hubble tension  can be an indicator of certain, still unnoticed, genuine differences in the descriptions of the late and early Universe \citep{R,R1,Val,Dai1,Dai2}. Namely, the discrepancy between the Hubble constant values obtained from local surveys and global, cosmic microwave background data can imply the need to also consider nonidentical mechanisms of cosmic structure formation at local and global scales (e.g., \citealt{photon,Bo}).

The formation of the large-scale matter distribution of the cosmic voids and web, filaments, and clusters of
galaxies are considered a result of the evolution of initial density fluctuations \citep{Peeb}. Zeldovich pancake
theory \citep{Z,Arn,SZ,SS} provides deep insights into certain features of cosmic structures.

Among various approaches to addressing the Hubble tension is the prediction of two Hubble flows, local and global ones, which are described by two similar sets of equations but nonidentical Hubble constants. The local flow is based on the following principal concept: \citet{G1} proved (as a theorem) that the general function for the force satisfying the identity of sphere-point gravity  has the form
\begin{equation}
\mathbf{F}(r) = \left(-\frac{A}{r^2} + \Lambda r\right)\hat{\mathbf{r}}\ .
\end{equation}   
It is remarkable that this function satisfies the first statement of Newton's shell theorem  (i.e., the sphere-point identity) but not the second part, namely, the force-free field inside a shell (for details, see \citealt{G2,GS2}).  Equation (1), as weak field general relativity (GR), allows the dynamics of groups and clusters of galaxies to be described \citep{GS3,GS4}. Observational indications include the influence of halos on the properties of spirals in galaxies (e.g., \citealt{Kr}), which supports the notion of a non-force-free field inside a shell predicted by Eq. (1).

Due to the second term in Eq. (1), the cosmological constant, $\Lambda$, emerges in nonrelativistic cosmologies, namely those of \citet{MM} and \citet{Z81}.  
As a result, one has two equations that look similar but have drastically different contents \citep{GS7,GS8}:
 \begin{equation}
H_{local}^2 = \frac{8 \pi G \rho_{local}}{3} + \frac{\Lambda c^2}{3},
\end{equation}
\begin{equation}
H_{global}^2 = \frac{8 \pi G \rho_{global}}{3} + \frac{\Lambda c^2}{3}.
\end{equation}
The first equation defines the nonrelativistic (non-GR) local flow determined by the repulsion of the $\Lambda$ term in Eq. (1) and, hence, with a {local} Hubble parameter. The second equation is the Friedmann equation for the Friedmann–Lemaitre–Robertson–Walker  metric and flat geometry, with the {global} Hubble parameter. As shown in \citet{GS8}, an analysis of these equations that takes the difference in the content of the matter mean densities into account -- local and global ones ($\rho_{local}$ and $\rho_{global}$) --  explains the quantitative discrepancy between the local and global values of the Hubble parameter and thus the Hubble tension.
   
In this context, our present analysis assumes a principal difference in the treatments of the late (local) and early (global) Universe and hence the need of different techniques to address the formation of structures on local and global scales. Namely, while the evolution of primordial fluctuations is considered to be responsible for the global matter distribution, the role of the self-consistent gravitational interaction can play an important role in the formation of local structures. Hence, we continue the kinetic Vlasov analysis 
\citep{web} of the formation of the cosmic voids and walls.     

We show the emergence of structures, two-dimensional walls and voids, based on an analysis of the properties of solutions of kinetic Vlasov--Poisson system equations, where the fundamental role has the potential of Eq. (1) with the repulsive cosmological constant term. Particularly, using the methods of bifurcation theory and analyzing solutions of integral equations, we show the possibility of multiply connected two-dimensional gravitating structures appearing. The results predict the dependence of void scales on the local physical parameters.

  \section{Vlasov--Poisson equations for modified gravitational interaction with a cosmological constant}

We began with consideration of the mechanism for the emergence of cosmological structures within
a nonrelativistic McCrea-Milne model of expansion of the Universe
with the modified gravitational interaction of Eq. (1) \citep{G1}. We considered a sphere of mass $M(R)$ that contains particles of the same mass,
$m$, such that the mass of particles in the ball $M = \int m f({\bf x},{\bf v},t)\:dm d{\bf v}$ (including
in the case of a uniform distribution, $M=4\pi \rho R^3/3$), and the law of conservation of energy in the form
\begin{equation}
\frac{1}{2}\bigg(\frac{d R}{dt}\bigg)^2 - \frac{G M}{R} -
\frac{c^2\Lambda}{12}R^2 = E = {\rm{const}}.
\label{1}
\end{equation}
\noindent
For the Hubble parameter, $H$ (here defined as $dR/dt = HR$),
we get the equation
\begin{equation}
\frac{dH}{dt}+H^2 +\frac{4\pi G \rho}{3}-\frac{1}{6}c^2\Lambda =0.
\label{2}
\end{equation}
We assumed the values $R(t=t_0)=R_0$, $H(t=t_0)=H_0$, and $\rho (t=t_0)=\rho_0$ at $t=t_0$.
We could thus define the value of the constant
$E=E_0= ({1}/{2})H_0^2R_0 -G \cdot \frac{4\pi}{3}\rho_0 R^2_0 -\frac{c^2}{12}\Lambda R_0^2$,
and the initial conditions for the differential equation for the evolution of a sphere of radius $R(t),$
\begin{equation}
\big(\frac{dR}{dt}\big)^2 =
\frac{a_1}{R}
- a_2 R^2
- a_3,
\end{equation}
$$
a_1 \equiv \frac{8\pi}{3} {G\rho_0 R_0^3},~~a_2 \equiv \frac{c^2\Lambda}{12}, ~~
a_3 \equiv \frac{8\pi}{3}G R^3_0 \big(    \rho_0 - \frac{3H_0^2}{8\pi G}-
\frac{3 c^2\Lambda}{8\pi G R_0}        \big).
\label{3}
$$
Solutions of  Eq. (\ref{3}) can be expressed in terms of degenerate elliptic Legendre integrals
of the third kind (see, for example, \citealt{Bat}):
$$
R(t) = I^{*}_3(-a_2,0,-a_3/6,a_1/4,0).
$$
At $\Lambda \equiv 0,$ one has the classical 
open (infinitely expanding) and closed Friedmannian models.
In this case, the Cauchy data for the considered ordinary differential
equations are usually taken as $R(t=0)=0$, $R_t'(t=0)=+\infty$.
It should be noted that the case of the nonzero cosmological constant $\Lambda$ 
does not lead to a fundamental change in the structure of solutions to the Cauchy problem (Eq.\ (\ref{3}); see \citealt{VFC1,VFC2}).

The McCrea-Milne model with the modified potential of Eq.(1) is stable
with respect to a change in the initial data, which is due to
the self-similarity of its solutions and the theorem in \cite{G1} on the general form of the potential
for a spherical region.
Thus, to study the structure formation in the local Universe,
one can use the Vlasov--Poisson equations without taking the GR effects  into account.
Obviously, the presence of an additional repulsive $\Lambda$ term on the right side of the Poisson equation is not due to the GR metric but to Eq. (1).

We are interested in analyzing the properties of deviations from the state of local equilibrium of
systems governed by Vlasov-Poisson equations. As the uniform distribution does not imply equilibrium in the general case in a system of massive
particles with an additional oscillatory interaction of geometric genesis (for the total super-harmonic potential), one can
study the evolution of matter in the vicinity of particle clusters as a result
of the evolution of cosmological fluctuations. Then, one can introduce the concept of relative equilibrium, where the distribution of particles of a given equilibrium state satisfies the principle of maximum entropy and, for example, have the property of averaging over
a set of hydrodynamic macro-parameters (see \citealt{VFC1,VFC2}).

The system of Vlasov--Poisson equations that describes the dynamics of a many-particle system
for a $d$-dimensional  case $(d=2,3)$     for  particles of equal masses, $m$, within the representation of a self-consistent gravitational field is written in the following form \citep{V1,V2},
\begin{equation}
\frac{\partial F({\bf x},{\bf v},t)}{\partial t} +
{\rm{div}}_{\bf x}({\bf v}F) - \frac{\partial F}{\partial {\bf v}}\nabla_{\bf x}\Phi(F)=0,
\end{equation}
$$
\Phi(F) \equiv
\int_{\omega_{\bf x'}}\int_{\omega_{\bf v'}}  {\mathfrak Y}_d
(|{\bf x}-{\bf x}'|)  F({\bf x}',{\bf v}',t_*)\:d{\bf x}'d{\bf v}',
\label{4}
$$
\noindent
where ${\mathfrak Y}_d({\bf x}-{\bf x}') \sim Y_d^{(1)}(|{\bf x}-{\bf x}'|)+Y_d^{(2)}\cdot |{\bf x}-{\bf x}'|^2$ (the function corresponding to the attraction between particles) is
$Y_d^{(1)}({\bf x}-{\bf x}') \in L_1^{loc}\bigcap C^2(\bar{\omega})$ and is integrable (and has a proper smoothness outside zero) in some bounded domain $\omega_d={\omega}_{\bf x} \subset {\mathbb R}^d$ function.
 The $Y_d^{(2)}\in {\mathbb{R}}^1$ is a function that depends on the cosmological constant, $\Lambda$.
We have assumed that, from the point of view of an external observer, the cosmological expansion  affects the substructures of the quasi-two-dimensional physical
structures in the same way as in the three-dimensional case.

It can be shown (see Sect. 3) that, for an arbitrary finite time $t_* \in {\mathbb R}^1_{+}$, the modified Poisson equation is
\begin{equation}
{\Delta}_{\bf x}^{(d)}\Phi(F)\big|_{t=t_*, \forall{t_*}\in
 {\mathbb R}^1_{+}}= \frac{2\pi^{d/2}}{\Gamma (d/2)} G \rho ({\bf x})-C_1,
\end{equation}
$$
\rho ({\bf x})\sim m \int F({\bf x},{\bf v},t_*)\:d{\bf v}
~~~(G_3 \equiv G,~~C_1=C_1(\Lambda)).
\label{5}
$$
Equation (\ref{5}) takes the form of the so--called (inhomogeneous) Liouville--Gelfand equation
\citep[LGE;][]{Gelfand}  if we take the density
distribution of particles equal to that in the case
of the Maxwell--Boltzmann integration
distribution function.\ Hereinafter we do not take  into account the dependence
of the equilibrium distribution function
on integrals of motion other than energy:
$F=F_0({\bf x},{\bf v},t_*)\propto \Omega
\big(\varepsilon[{\bf v}(t_*);\Phi ({\bf x})]\big|_{ext} \big)$ in terms of velocities
(a regular branch of $\Phi$ is chosen).
In this case, in the energy neighborhood of the local extremum, one has
\begin{equation}
d\Omega/d\varepsilon\big|_{r \to r_c}\approx 0,~~~~r_c = \big(G m/(3\Lambda c^2)\big)^{1/3}
.\end{equation}
\noindent

In this case, the force term in the Vlasov equation is either vanishing or
possesses a turning point on the phase plane of the evolution of the system, and
we get a relative equilibrium  (or quasi-stationary) state as a solution of
 the Vlasov--Poisson equations system.  It is characterized by the average quasi-equilibrium statistical temperature of the system,
$T$, and the potential gauge or density, $\rho_0=A\exp(-\Phi (0)/T),$ or
the spatial distribution in a point\:${\bf x}=0$, as well as by the selected energy level of particles on the regular branch of the potential. For the sake of simplicity, we provide Eq.\ 10, an explicit form of the LGE for the two-dimensional case:
\begin{equation}
-\Delta^{(2)}U=\lambda \exp(U) -c^2\Lambda/T,~~~U=-\Phi/T,
\end{equation}
$$
\lambda = 2 G mN \rho_0/T,~~~T\equiv -
\frac{\partial \Phi/\partial x_j}{\partial \rho/\partial x_j}\big|_{j=1,2}.
$$
From a mathematical point of view,  the potentials (solutions of the LGE) correspond to the continual
matter density distribution (in the $\omega \subset {\mathbb R}^d$ calculation area with given conditions on the $\partial \omega$ boundary).
Obviously, this distribution is definitely idealized from a physical point of view.
However, based on Theorem 1.3 in \citet{Esp} and \citet{Bre}, it can be argued
that, when the internal parameters of the function $\lambda (\rho_0,T)$ are changed in
a limited area of space, there is a system of
discrete masses, ${\bf q}_i\big|_{i=\overline{1,k}}$ (defined in terms of singular points of
Robin functions), whose action on a distant particle is equivalent to the given continuous distribution of matter:
\begin{equation}
\frac{\displaystyle \lambda \exp(U_\lambda)}{\displaystyle
\int_\omega \exp(U_\lambda)d{\bf x} }\rightharpoonup 8\pi
\sum^k_{i=1}\delta_{{\bf q}_i}.
\end{equation}
Since there is scale invariance here, we can see that the replacement of
discrete particles on a continuum distribution
is valid both for a local single cluster (for example, understood as an elementary
substructure) and for the system of two-dimensional walls. The gravitational potential in the
neighborhood of the extremum point is close to constant, and the influence of the self-consistent
fields on particle motion
is absent; therefore, the inertial motion of particles leads to relatively long-duration formations whose local topology is not distorted.

We have assumed that the variation in the action
over the gravitational field (resulting in the Poisson equation) and variation over particles
(leading to the equations of motion along geodesics, and after the introduction of the distribution function
particles to the Liouville equation) are independent operations. Thus, we can separate in the first approximation
the particle distribution evolution equation and equations for self-consistent gravitational
fields, assuming
that particles move in a given field and that changes in the particle distribution function
occur in accordance with the solution of the Liouville equations \citep{Vedkni}.
Thus, we can study the equation for the potential separately, although
it is essentially nonlinear.\ To study it, one needs to use
methods of branching theory for solutions of nonlinear equations. However, important conclusions can be drawn 
even from the linearized version of the equation for the gravitational potential, which we demonstrate below.

  \section{Equation for the gravitational potential in integral representation }

We considered a system of $N$ point bodies with equal masses, $m$, interacting via the modified gravitational potential of Eq. (1).
The Hamiltonian of the system has the form
\begin{equation}
({\mathfrak{H}}_N)_\epsilon=\sum_{j}^N\bigg(
\frac{{\bf p}^2_j}{2m} + {\sum_{i}}' \Phi_\epsilon^{(2)} (|{\bf x}_i-{\bf x}_j|) + m\Phi_{ext} ({\bf x}_j)+B(\partial{\omega},{\bf x}_j)
\bigg),
\label{6}
\end{equation}
\noindent
where $\Phi_{ext}({\bf x}_j)$ is the potential of the external gravitational field at a point (${\bf x}_j$), and $B(\partial{\omega},{\ bf x}_j)$ is the potential energy of the interaction of particles with boundaries,
which includes the reflection of particles from the boundary (periodicity of the boundaries) and the impact due to a change in the local temperature at the boundary of near-equilibrium particles inside the $\omega$ region
(see \citealt{Kis1,Kis2}).
The index $\epsilon$ of the two-particle potential $\Phi^{(2)} \big(r({\bf x},{\bf x}') \big)\sim c_1/|{\bf x}-{ \bf x}'|+c_2 |{\bf x}-{\bf x}'|^2$ implies the regularization of
a weakly polar function with a small argument; in other words, under the integral we replaced (e.g., in the three-dimensional case)  the function $\Phi^{(2)} (r)$ with the function
\begin{equation}
\Phi_\epsilon^{(2)}  \big(   r({\bf x},{\bf x}')  \big)=
\end{equation}
$$
\begin{cases}
( c_1 r({\bf x},{\bf x}') )^{-1} + c_2 r^2({\bf x},{\bf x}'),\,\,        & r({\bf x},{\bf x}') \geq \epsilon;\\
(2\epsilon)^{-1}\cdot \big(  3-  |r({\bf x},{\bf x}')|^2/\epsilon^2   \big) + c_2 r^2({\bf x},{\bf x}'),\,\,      & r({\bf x},{\bf x}')< \epsilon.
\end{cases}
$$
\noindent
In this case, the volume integral potential with a summable bounded density of particles is a continuous function up to the boundaries of the system.
In the absence of significant gradients of macro-characteristics, the particles of the system can be described using the formalism of the canonical ensemble,
characterized by the corresponding (near-)equilibrium
density in the phase $6N$-dimensional space:
\begin{equation}
\varrho_N =(N! \Omega_\epsilon)^{-1}\exp(-{\mathfrak{H}}_N/T),
\end{equation}
\begin{equation}
\Omega_\epsilon= (N!)^{-1}\int\exp(-({\mathfrak{H}}_N)_\epsilon/T)\prod_j d{\bf x}_jd{\bf p}_j.
\end{equation}
The canonical probability measure of particles in the configuration subspace has the form
\begin{equation}
\mu_N (T)=\Theta^{-1}_\epsilon \exp\big(
-\frac{1}{2}\sum_{j<i} \Phi_\epsilon (|{\bf x}_j-{\bf x}_i|)/T
\big)d\omega^N,
\end{equation}
\begin{equation}
d\omega^N \equiv \prod_k^N \exp \big( -m\Phi({\bf x}_k) /T \big)d{\bf x }_k,~~
\end{equation}
\begin{equation}
\Theta_\epsilon = \int\exp_{\omega}\big(
-\frac{1}{2}\sum_{j,i; \:j \neq i} \Phi_\epsilon (|{\bf x}_j-{\bf x}_i|)/T
\big)d\omega^N,
\end{equation}
\noindent
and the divergence of the configuration integral $\Theta_\epsilon\to \infty$ at $\epsilon\to 0$ yields $|{\bf x}_j-{\bf x}_i|\to 0$. Separating two particles from the $N$-particle
ensemble (${\bf x}_N \equiv {\bf x}$, ${\bf x}_{N-1} \equiv {\bf x}'$), we can consider the
density of the probability measure ${\mathfrak P}_N ({\bf x},{\bf x}',\omega_{N-2};\epsilon)$:
\begin{equation}
{\mathfrak P}_N ({\bf x},{\bf x}',\omega_{N-2};\epsilon) = 
\end{equation}
$$
\Theta^{-1}_\epsilon (N,T)\exp\big(
-\Phi_\epsilon (|{\bf x}-{\bf x}'|)
\big)\mathcal{P}({\bf x},{\bf x}',\omega_{N-2};\epsilon),
$$
\noindent
where $\mathcal{P}(...)$ is a positive function that includes the interaction potentials of other particles, and ${\mathfrak P}_N ({\bf x},{\bf x}',\omega_{ N-2};\epsilon\to 0)\to 0$
(${\bf x}\neq{\bf x}'$).
So, $w^*-{{\lim}}_{\epsilon\to 0}{\mathfrak P}_N ({\bf x},{\bf x}',\omega_{N-2} ;\epsilon)={\mathfrak P}_{N-2}({\bf r}) \delta ({\bf x}-{\bf x}')$. Then, we can
consider the canonical ensemble in the mean field limit for de-singularized potentials. According to the Hewitt--Savage representation theorem on the permutation-invariant probability measure \citep{HS},
any such (admissible) measure, $\mu$, of the configuration subspace of the phase space of the many-particle system under consideration can be represented as
an integral over $d\varrho =\rho ({\bf x})d^3x$, $\rho ({\bf x})$ and can be considered in the classical density form. The free energy functional for our
system in this case has the form
\begin{equation}
{\mathcal F} (\rho) =\frac{1}{2}\int_\omega\int_{\omega'} \rho ({\bf x})\rho ({\bf x}') N \Phi_\epsilon (|{\bf x}-{\bf x}'|)d{\bf x}d{\bf x}' +
\end{equation}
$$
\int_\omega
\frac{1}{2}\rho ({\bf x}) \zeta ({\bf x}) d{\bf x}
+\int_\omega \rho({\bf x})\ln\big( \rho({\bf x}) \big)d{\bf x},
$$
\begin{equation}
\zeta ({\bf x}) = m \Phi_{ext}({\bf x}) + B (\partial \omega, {\bf x}).
\end{equation}
The minimization of ${\mathcal F} (\rho)$ occurs with the functions
\begin{equation}
\rho ({\bf x}) = {\mathfrak{R}}^{-1}\cdot \exp\bigg(
T^{-1}\big(
\int_{\omega'} \rho ({\bf x}')N\Phi_\epsilon (|{\bf x}-{\bf x}'|)d{\bf x}' +\zeta ({ \bf x})
\big)
\bigg),
\label{7}
\end{equation}
\begin{equation}
{\mathfrak{R}} \equiv
\int_{\omega''}\exp\bigg(
T^{-1}\big(
\int_{\omega'} \rho ({\bf x}')N\Phi_\epsilon (|{\bf x}''-{\bf x}'|)d{\bf x}' +\zeta ({\bf x}'')
\big)
\bigg)d{\bf x}''.
\end{equation}
Since the most probable state of a thermodynamic system corresponds to the minimum free energy,
the solution of Eq. (23) corresponds to the equilibrium density distribution in the multi-particle system.

If we recall the original Vlasov--Poisson equation, Eq. (\ref{4}), the simplest stationary solution of Eq. (23) is the function $f({\bf x},{\bf v})=\widehat{f}(\chi)$,
where $\chi = m{\bf v}^2/2 + \Phi ({\bf x})$ (energy substitution) and $\widehat{f}(\chi) \in C^2 ({\mathbb R})$.The Poisson equation
when choosing $\widehat{f}(\chi)=c_1\exp(-\chi/T)$ (the Maxwell--Boltzmann equilibrium distribution known from collisional theory)
can be represented as
\begin{equation}
\Delta^{(d)} \Phi= G\varkappa_d (T) \exp \big( -\Phi/T \big) -c^2\Lambda,
\label{8}
\end{equation}
\begin{equation}
\varkappa_d (T) =
c_0 |s_d|^2 \int_{{\mathbb R}^1_{+}} \exp \big(-\eta^2/(2T) \big)\eta^{d-1}d\eta,
\end{equation}
\noindent
where $|s_d|$ is the area of the $d$-dimensional sphere of unit radius. In this case, the potential $\Phi ({\bf x})$ refers to the averaged self-consistent
field determined by integrating into formula (\ref{7}) the local densities of particles with a nucleus in the form of an inter-particle potential, $\Phi_\epsilon ({\bf x}-{\bf x}')$. Accordingly, the density of particles, as noted in Sect. 2, can be considered as a continuous argument function, and with it we can pass to the weak topology of isolated points.
The equivalence of temperatures for the canonical and micro-canonical descriptions, generally speaking, is not too obvious, but significant problems can only arise for
strongly nonequilibrium systems with negative thermodynamic temperatures.  The influence of system boundaries (Eq. \ref{7}),
in turn, is extremely important. We will deal with this issue in the future, using a more appropriate formalism.

The form of the LGE for the gravitational potential in cosmological systems is essentially local (as it is a differential equation), and it is difficult to use it to describe global processes in a self-consistent field in order to analyze the evolution of the entire system of particles
(substructures). When studying nonlocal effects,
it is expedient to consider the integral version of the equation for the potential. However, a natural limitation for the use of this form in analytical and numerical calculations is the obvious need to take into account the boundaries of the region that contains the system of particles with
% (in which region the distribution of $(N-1)$ particles of the system takes place)
  regards to the value of the potential at a given point (i.e., the explicit form and effects of the term $\zeta ({\bf x})$ in Eq. (\ref{7})).
Of course, this is due to the fact that, for a system of particles interacting according to the law of Eq.(1), the zero Dirichlet condition, $\Phi ({\bf x}\big|_{\partial \omega})=0$,  has to apply to the boundary of the region under study.\ This condition is subtle, for example, for the general, though unrealistic, case of an asymmetric  many-particle system.

We considered the following physical problem regarding the structure of the local (late) Universe where there are high density regions, consequences of cosmological fluctuations at earlier phases: can the gravitational interaction between these $\omega_j$ regions (each of which contains, e.g., a hierarchy of substructures of various sizes)
or in the vicinity of each region form a (quasi-)stationary ordered structure?
The essential difference compared to a charged plasma with Debye screening, which allows
pseudo-homogeneous equilibrium particle distributions, is that the cosmological substructures
evolving from primary perturbations of density
  macroscale objects are spherically symmetric ($d=3$) or radially symmetric
  (for $d=2$), according to the Gidas--Ni--Nirenberg theorem \citep{Gi}.
  In accordance with the structure of the inter-particle potential, ${\mathfrak Y}_d$, in the neighborhood
  spheres with increased density, a radially symmetric layer arises, in
  which the attraction to the center of the sphere interplays with a
  semi-infinite spherical layer with a prevailing repulsion. For a system of two
  spheres that contain matter of increased density, there is competition between those two centers,
  leading either to the destruction of one of them or, in connection with the cosmological
  expansion, to the emergence of repulsion zones between them.

Two approaches can be considered when searching for the gravitational potential in the vicinity of a region that contains a system of interacting particles. 

The first is a solution of the equation for the potential (inhomogeneous LGE) under given Dirichlet conditions on some a priori defined boundary of the region, $\partial \omega_j$; this boundary should include only one local overdensity. In this case, one can consider both the internal and external Dirichlet problems, which correspond to attempts at establishing the values of the potential (and, hence, the density of particles) inside and outside the $\partial \omega_j$ boundary shell, respectively. Influence
potentials of neighboring regions can be neglected under certain conditions. The data at the boundaries must be consistent with the resulting solution (i.e., they must be a continuity of the solution).

A second, similar situation can also be considered for the Neumann problem; however, it is possible to define
the boundary of the $\partial \omega_j$ domain due to the presence of a maximum of the interaction potential (at this point, the derivative of the potential vanishes).
However, in such a formulation, a question naturally arises regarding the legitimacy of describing the dynamics of the region of the particles  only as the zone of attraction of the potential.

The physical aspects of this problem obviously determine the way additional conditions for the nonlinear potential equation are set.
We restricted ourselves to using the Dirichlet conditions (due to the shell theorem on the equivalence of the gravitational field of a sphere and a point at its center).
Thus, we are interested in whether inside (or outside) the fixed region, $\omega$, for a given
value of the potential at the boundary, there are secondary solutions of the LGE, Eq. (\ref{8}), 
that possess a (quasi-)periodicity property.

The solution of the Dirichlet problem for the Poisson equation, according to \cite{Gil} and \cite{Sau}, can be represented through the Green's elliptic operator and the value of the
potential at the borders. Explicitly, this solution is
\begin{equation}
\Phi ({\bf x}) = -\int_{\omega}{\mathcal G}({\bf x},{\bf x}')\bigg( G \rho ({\bf x}' ) -\frac{c^2\Lambda}{4\pi} \bigg)\:d{\bf x}'
-
\end{equation}
$$
\frac{1}{4\pi} \int_{\partial \omega'} \Phi|_{\partial \omega}{\bf n}\cdot
\frac{\partial}{\partial {\bf x}'}{\mathcal G}({\bf x},{\bf x}')|_{\partial \omega}dS',
\label{9}
$$
where ${\mathcal G}({\bf x},{\bf x}')$ is Green's function of the Dirichlet problem for the inhomogeneous Poisson equation.
If we choose a ball of radius ${\mathcal{R}}$ ($\omega=\{\tilde{\bf x};\, 0\leq|\tilde{\bf x}|\leq {\mathcal {R}}\}$),
then it is possible, if the distribution of particles is close to the isotropic one,
to assume the values of the potential at its boundary, in accordance with the theorem in \citet{G1}, as 
\begin{equation}
\Phi|_{\partial \omega}=-G M/{\mathcal{R}}-
c^2\Lambda {\mathcal{R}}^2/6, \, (M=Nm). 
\end{equation}
Replacing $\rho ({\bf x}) =(4\pi)^{-1}\varkappa_d (T) G \exp \big( -\Phi/T \big),$
and defining, according to the \cite{Now}, the Green's function for the considered Dirichlet domain,
\begin{equation}
{\mathcal G}({\bf x},{\bf x}') \equiv 4\pi \sum^\infty_{\ell =0} \sum_{m=-\ell}^\ell \frac{Y_{\ell m}^{*} (\theta',\varphi')  Y_{\ell m} (\theta,\varphi)}{2\ell + 1}
\frac{x_{<}^\ell x_{>}^\ell}{{\mathcal{R}}^{2\ell + 1}},
\end{equation}
$$
x_{<}={\rm{min}}(|{\bf x}|,|{\bf x}'|),~~x_{>}={\rm{max}}(|{\bf x}|,|{\bf x}'|),
$$
\noindent
we obtain a nonlinear integral equation for the potential $\Phi ({\bf x})$ (for $d=3$):
\begin{equation}
\Phi ({\bf x}) =- \varkappa_d (T) G \int_{\omega'}\bigg( \frac{1}{|{\bf x}-{\bf x}'|} -
{\mathcal G}({\bf x},{\bf x}')
\bigg) 
\end{equation}
$$
\exp \big( -\Phi ({\bf x}')/T \big) d{\bf x}' - \frac{c^2\Lambda}{6}{\bf x}^ 2 + C_0,~~
$$
\begin{equation}
C_0 = -\frac{G M}{{\mathcal{R}}} - \frac{c^2\Lambda {\mathcal{R}}^2}{6 }.
\end{equation}
After obvious transformations, this can be rewritten as an inhomogeneous Hammerstein-type equation for the dimensionless potential, $U_H$:
\begin{equation}
U_H ({\bf x}) =\widehat{\mathfrak{W}}\big(U_H ({\bf x})\big),~~~
\widehat{\mathfrak{W}}\big(U_H ({\bf x})\big)\equiv 
\end{equation}
$$
\lambda_H(T)
\int_{\omega'} \underbrace{\bigg( \frac{1}{|{\bf x}-{\bf x}'|}-{\mathcal G}({\bf x},{\bf x}') \bigg)}_{{\mathcal{K}}({\bf x},{\bf x}')}\exp\big(-U_H ({\bf x}') \big)
d{\bf x}' +
\label{10}
$$
$$
+\alpha (\Lambda,T) |{\bf x}|^2; 
$$
\begin{equation}
\lambda_H(T) \equiv \frac{-\varkappa_3 (T) G_3 }{T}\exp\big( - \frac{C_0}{T} \big),~~~ \alpha (\Lambda,T ) = -\frac{c^2\Lambda}{6T},
\end{equation}
$$
U_H = \frac{\Phi - C_0}{T}.
$$
It should be noted that, for a spherically symmetric density distribution,
\begin{equation}
\int_{\omega'} {\mathcal G}({\bf x},{\bf x}') \rho ({\bf x}')d {\bf x}'\to C_1 (={\rm{const}}).
\end{equation}

  \section{Formation of voids}

Equation (\ref{10}) contains information about the behavior of the considered cosmological, non-GR dynamics of a many-particle system with the gravitational interaction of Eq. (1).
First of all, we should point out
the obvious absence of a solution for this equation for $d=3$ in the entire area of study
(and in the whole space, except for two-dimensional surface isograves) in the form of a constant potential, in contrast to the classical Newtonian case of only attractive gravity.
So, in this case, the global
uniform distribution of matter, as a condition for matching the modified Poisson equation and the Vlasov equation, can only be considered as a certain approximation; a more comprehensive analysis is needed for more general cases.

We considered a linearized (near the point $U_H(0)=U_0$) version of Eq. (\ref{10}) in the homogeneous and inhomogeneous cases:
\begin{equation}
U^\dag =\widehat{\mathfrak{W}}_0'\big(U^\dag\big),~~~U^\ddag =\widehat{\mathfrak{W}}_0'\big(U ^\ddag\big)+\alpha |{\bf x}|^2.
\label{11}
\end{equation}
Operator $\widehat{I}-\widehat{\mathfrak{W}}_0'$, where $\widehat{\mathfrak{W}}_0'\equiv \widehat{\mathfrak{W}}'[U_0]$ is a Frechet  derivative of  the Hammerstein integral operator
on the right side of the first equation, belongs to
the class of zero-index Noetherian operators with a weak singularity. Since Green's function is symmetric to its coordinates,
for matter distributions that slightly deviate from spherically symmetry, $\widehat{\mathfrak{W}}_0'$ can also be considered self-adjoint.
For a small deviation in the amplitude of $\delta U^\dag$ from the selected solution, one can apply the well-known mathematical apparatus for the analysis of Fredholm operators to the homogeneous linear equation $\delta U^\dag - \lambda \int_{\omega'} {\mathcal{K}}({\bf x}-{\bf x}')\exp(-U^\dag_0 )\delta U^\dag d{\bf x}'=0$. To simplify the calculations without
losing generality, we can take $U_0^\dag=0$ and look for periodic solutions of the last equation in the form of an expansion in terms of
eigenfunctions ${\mathfrak{b}}_j =c_\omega \exp(i{\bf{q}}{\bf x})$ of the kernel
${\mathcal{K}}$ in $\omega \subseteq {\mathbb R}^3$) in the form
$\delta U^\dag = \sum_j {\mathfrak{a}}_j (c_\omega)\exp(i{\bf{q}}{\bf x})$ ($q_{\ell=1, 2,3}=2\pi/d$; the cubic case can be generalized to $d_{\ell}\neq d_{k}$).
Substituting this expression into the first population equation, Eq. (\ref{11}), we have
\begin{equation}
1= \lambda \int_{\omega'}{\mathcal{K}}({\bf x}-{\bf x}') \exp\big( -i{\bf q} ({\bf x} -{\bf x}') \big)\:d{|{\bf x}-{\bf x}'|}.
\label{12}
\end{equation}
Here we introduce a critical value of the parameter $\lambda = \lambda_c$, corresponding to the case $d \to \infty$, $q_{\ell}\equiv 0$. Using it, we can write
the criterion for the existence of periodic solutions for the linearized integral homogeneous Poisson equation:
\begin{equation}
\lambda = \bigg( \int_{\omega} {\mathcal{K}}(r)\frac{\sin(qr)}{qr} r^2\:dr d\theta d\phi \bigg)^ {-1}\ge \lambda_c \equiv
\end{equation}
$$
  \bigg( \int_{\omega} {\mathcal{K}}(r) r^2 \sin(\theta)\:drd\theta d\phi \bigg)^{-1}.
\label{13}
$$
Obviously, this criterion is suitable only for the case $\Lambda \equiv 0$, and only for the two-body interaction approximation. This implies  that the root, ${\bf q}$, of Eq. (\ref{12}) is unique and, thus, that the potential distribution will be purely periodic. As there are  several incommensurable roots of ${\bf q}_s$, the distribution of the potential will belong to the class of almost-periodic functions. The accounting of collective
 interactions of $N$ particles leads to a (homogeneous) equation of the form
  \begin{equation}
  \delta U^\dag ({\bf x})= \sum_{k=1,...,N}\int_{\omega_1}...\int_{\omega_k} {\mathcal{K}}_k ({\bf x},{\bf x}_1,...,{\bf x}_k)
        \end{equation}
        $$
  \big( \exp(-\delta U^\dag ({\bf x}_1))...\exp(-\delta U^\dag ({\bf x}_k))
  \big)\:d{\bf x}_1...d{\bf x}_k.
  $$
After linearizing this equation, we get (see \citealt{V1})
\begin{equation}
\delta U^\dag ({\bf x})= \sum_k \lambda_k \int_{\omega_1}...\int_{\omega_k}{\mathcal{K}}_k({\bf x},{\ bf x}_1,...,{\bf x}_k)
\end{equation}
$$
 \sum_\ell^k \delta U^\dag ({\bf x}_\ell)\prod_{s=1}^k
d{\bf x}_s.
$$
The corresponding criterion, criterion (4), for the occurrence of three-dimensional periodic solutions can be represented in the following form:
\begin{equation}
\int_\omega \sum_{k} \sum_{s}^k \lambda_k \frac{\sin(qr)}{qr}
\bigg(\int_{\omega_1}...\int_{\omega_{s-1}}\int_{\omega_{s+1}}...\int_{\omega_k}
\end{equation}
$$
{\mathcal{K}} _s({\bf x},{\bf x}_1,...,{\bf x}_k)\frac{\prod_{n=1}^k d{\bf x}_n}{d{\bf x}_s}\big)
r^2 \sin(\theta)\:dr d\theta d\phi =1.
\label{14}
$$
Therefore, the accounting of the cluster interactions in a many-particle system is consistent with the accounting of two-particle interactions: the appearance of a periodic structure
potential and of the density of matter in a linear homogeneous approximation occurs abruptly upon reaching
a certain, quasi-equilibrium temperature in the system. The periods of the structures are determined from conditions (\ref{12}) and (\ref{14}).

For an inhomogeneous linear equation (the second half of Eq. (\ref{11})), it is possible, using the Hilbert--Schmidt theorem \citep{Tri}, to obtain an explicit form of the (unique)
solutions in the form of a resolvent series, as uniformly and absolutely convergent Fourier series in the eigenfunctions of the kernel ${\mathcal{K}}$, for $\lambda \neq \lambda_\ell$ ($\ell=1,2,... $):
\begin{equation}
\delta U^{\ddag} ({\bf x})= \lambda \int_\omega \sum_{\ell =1}^\infty \frac{{\mathfrak{b}}_\ell ({\bf x}) {\mathfrak{b}}_\ell ({\bf x}')}{\lambda_\ell - \lambda}
\alpha |{\bf x}'|^2\:d{\bf x}'
  +\alpha |{\bf x}|^2,
\end{equation}
\noindent
where $\lambda_\ell$ are the characteristic numbers of the homogeneous equation (i.e., they correspond to the eigenfunctions ${\mathfrak{b}}_\ell ({\bf x}))$.

Thus, the periodicity of the structure, when $\Lambda$ repulsion is taken into account,  degenerates (in the simplest case) into a composition of  sinusoidal functions and  growing
branches of  parabolas, which leads to a smoothing of the periodicity and the dominance of repulsion at a certain distance,
in the line of interactions between two formed inhomogeneities
in a pseudo-homogeneous distribution of matter.  This process can be considered a pair interaction and a mechanism for the formation of voids  in an initially uniformly distributed material continuum. The anisotropy of particle momentum directions in the channels between $I$ and $II$ inhomogeneities forms
not a cubic lattice
($d_{1,2,3}=d$) but quasi-one-dimensional layers ($d_3 \ll d_{1,2}$). Their velocity profile in these channels is decelerated for initially fast particles and thus is synchronized
as a uniform distribution in velocity space,
due to repulsion in the far zone of
the second component of the macro-system (i.e., the $II$ inhomogeneity at the other end of the channel). It can be assumed that a significant part of the density
inhomogeneity transforms into a flat structure corresponding to the first
potential minimum and that this situation is mirrored in the channel  of pair interaction. The repulsion acts as an external force,
and from the $II$ side leads to the quasi-stationarity of the wall formed by the $I$ inhomogeneity and vice versa. The $\Lambda$ repulsion leads to initially attracting massive inhomogeneities
appearing at distances at which the repulsion becomes more significant due to the influence of the cosmological term.

As shown in \citet{GS7,GS8}, based on an analysis of observational data, the repulsive term in Eq. (1) can dominate the
attraction term depending on the mean matter density of certain galaxy clusters.
Then, based on the above analysis, one can conclude that, although the considered mechanism of the void
formation is common – the mutual interplay of the self-consistent
gravitational attraction and of the repulsion due to the
cosmological constant term – the voids can possess different mean scales that are
determined by the local conditions, the density, and the initial momenta of particle flows.
Thus, our proposed method for analyzing the solutions to the Poisson equation is able to explain the presence of voids of various scales.

The solution of the nonlinear Hammerstein equation for the potential differs significantly from the solution of the Fredholm equations. If first we turn to the case $\Lambda \equiv 0$ (homogeneity of the equation), we can immediately see that Eq. (\ref{10}) has a constant solution, $U_H = U_H^{(0)}={\rm{const}}$.
From a physical point of view, this means that we are considering a medium that consists of particles, the interaction between which corresponds to
only their mutual attraction, in which there are no primary perturbations. Obviously, this is an extremely unstable system.
But even without an external influence leading to the local coalescence of particles, when the parameter $\lambda$ reaches a certain value, solutions of a new type arise, branching off from the constant.

We denote $\lambda_H = \lambda_H^{(0)} +\eta$, $U_H ({\bf x})= U^{(0)}_H + u ({\bf x})$. Then, if we expand  the exponent   in the integrand
expression in a Taylor series, following \cite{Bra}, we get
\begin{equation}
u ({\bf x}) - \lambda_H^{(0)} \exp(-U^{(0)}_H) \int_{\omega'} {\mathcal{K}}({\bf x} -{\bf x}') u ({\bf x}') d{\bf x}' =
\label{15}
\end{equation}
$$
\eta \exp(-U^{(0)}_H) \int_{\omega'} {\mathcal{K}}({\bf x}-{\bf x}') d{\bf x}' -
\eta \exp(-U^{(0)}_H) 
$$
$$
\int_{\omega'} {\mathcal{K}}({\bf x}-{\bf x}') u ({\bf x} ') d{\bf x}' +
$$
$$
+ (\eta + \lambda_H^{(0)})\exp(-U^{(0)}_H)\sum_{j=2}^\infty (-1)^j (j!)^{- 1}\int_{\omega'}
{\mathcal{K}}({\bf x}-{\bf x}') \big( u ({\bf x}')\big)^j d{\bf x}'.
$$
Substituting $u({\bf x})= \sum_{s=1}^{\infty}(\eta \exp(-U^{(0)}_H))^{s/2} {\mathcal X}({\bf x})$ (here ${\mathcal X}({\bf x})$ are unknown
functions to be defined), we obtain a sequence of linking equations with a linear left-hand side of the form 
\begin{equation}
\widehat{\mathfrak{N}}({\mathcal D}_1)=\int_{\omega'} {\mathcal{K}}({\bf x}-{\bf x}')d{\bf x}'=0,~
\widehat{\mathfrak{N}}({\mathcal D}_2)=\int_{\omega'} {\mathcal{K}}({\bf x}-{\bf x}')
\end{equation}
$$
\bigg(
1+ \lambda_H^{(0)}\exp(-U^{(0)}_H) \frac{({\mathcal D}_1({\bf x}'))^2}{2!}
       \bigg)d{\bf x}',...,
$$
\begin{equation}
\widehat{\mathfrak{N}}({\mathcal D}_j) \equiv {\mathcal D}_j+\lambda^{(0)}\exp(-U^{(0)})\int {\mathcal {K}}({\bf x}-{\bf x}'){\mathcal D}_j({\bf x}')d{\bf x}'
\end{equation}
(the operator
associated with the homogeneous Fredholm equation). We have already obtained the form of the first term of a series of successive approximations: ${\mathcal D}_1 ({\bf x})=c_\omega\cdot \sin({\bf q}{\bf x})$.
The equation $\widehat{\mathfrak{N}}({\mathcal D}_1)=0$ has periodic solutions under the criterion $\lambda^{(0)}\exp(-U^{(0)} \Omega({\bf q})+1=0$,
  $\Omega({\bf q}) \equiv 4\pi \int {\mathcal{K}}(r)\big( \sin(qr)/(qr) \big)r^2dr$). In accordance with \cite{V1}, we successively solved the above equations and obtain a solution for $u({\bf x}):$
 \begin{equation}
  u({\bf x}) =\sum_{m=0} \big(\eta\exp(-U^{(0)}) \big)^{m+1/2} \bigg( \sum_{ \ell =1}^{m} {}_{(I)}C^{(2m+1)}_{2\ell + 1} 
        \end{equation}
        $$
        \sin\big((2\ell+1){\bf q }{\bf x}\big) +
  {}_{(III)}C_{2m+1} \sin\big( {\bf q}{\bf x} \big)
  \big) +
$$
$$
+\sum_{m=1}^\infty \big(\eta\exp(-U^{(0)}) \big)^m \bigg( \sum_{\ell =1}^{m} {} _{(II)}C^{(2m)}_{2\ell}\cos\big(2m {\bf q}{\bf x} \big) +
$$
$$
{}_{(III)}C_{2m} \sin\big( {\bf q}{\bf x} \big) \bigg).
$$
The resulting series, subject to its convergence, is a
periodic function with $T=T(|{\bf q}|)$. Since all coefficients ${}_{(i)}C_{...}$ can be obtained explicitly by sequential calculation,
we are able to determine the numerical values of the amplitudes of harmonics in a given Fourier series. Thus, one can obtain a ``fine structure'' of the potential, which means that the density of particles in the emerging structures branching off from the cosmological solution with a constant density parameter, $\eta$, 
is related to the difference between the temperature of the medium and the critical temperature corresponding to the critical parameter, $\lambda_c$.

As already mentioned, this technique requires knowledge of the ``primary'' solution to the homogeneous Hammerstein equation for the potential.
Using the methods of the perturbation theory of Fredholm operators (e.g., \citealt{Fre1,Fre2}),
  one can construct a solution for the inhomogeneous Hammerstein equation based on the results obtained above.
  If we denote ${\widehat{I}}_{+}\equiv \widehat{I}-\alpha |{\bf x}|^2$, then ${\widehat{I}}_{+}-\widehat {\mathfrak{W}}_0'$ is a Noetherian operator
  zero index, ${\rm dim}{\mathfrak{Y}}=1$, ${\mathfrak{Y}}\equiv{\rm Ker}({\widehat{I}}_{+}-\widehat {\mathfrak{W}}_0')$. The condition that
  ${\widehat{I}}_{+} - \widehat{\mathfrak{W}}_0'$ has a pseudo-inverse -- that is, a bounded inverse from the co-kernel complement ${\mathfrak{Y}}^{*\bot}$
  in some complement $\bar{\mathfrak{Y}}$ of the kernel ${\mathfrak{Y}}$ -- is equivalent to the statement about the existence of a unique element $v \in \bar{\mathfrak{Y}}$,
  such that $({\widehat{I}}_{+} - \widehat{\mathfrak{W}}_0')v=w$, $w \in {\mathfrak{Y}}^{*\bot }$. In this case,
  the nonlinear Hammerstein equation has a unique solution: $\big(\lambda_H (\varepsilon), u_H (\varepsilon) \big)$, where $\varepsilon$ is a small
  parameter associated with the temperature deviation from the critical one. The values $\lambda_H$ and $u_H$ are the limits of the sequences
  $\lambda^{\mu+1}(\varepsilon) = F_{+} (u^\mu(\varepsilon))$, $u^{\mu+1} (\varepsilon) =\varepsilon (\phi_0 + \varepsilon v^{\mu +1}(\varepsilon))$,
and  $v^{\mu+1}\in \bar{\mathfrak{Y}}$ ($u^0 (\varepsilon) =\varepsilon \phi_0$, $v^0 (\varepsilon)=0$, where $ \phi_0$ is an element
  ${\widehat{I}}_{+}-\widehat{\mathfrak{W}}_0'$ kernels). Thus, we have at our disposal a procedure for obtaining a (unique) solution of an inhomogeneous nonlinear
  equation for the potential. Near it, we can consider a certain neighborhood where new non-holomorphic branches of its solution can appear. However,
  it is already clear that the Hammerstein equation under study has a solution that is locally close in its construction to the inhomogeneous Fredholm equation. Further
  obtaining a solution constructed in a consistent way outside the uniqueness-domain branching solutions (replacing the constant solution for the homogeneous case)
   is practically equivalent to the homogeneous case.

Thus, it can be argued that the behavior of the potential in a many-particle system is capable of creating conditions for the emergence of two-dimensional structures, which can be associated with the walls of voids. The inclusion in the Poisson equation of an additional repulsive $\Lambda$ term from Eq. (1) plays a fundamental role here. Moreover,
using the formalism developed above, one can obtain the fine structure of the voids themselves, since the solutions of the Hammerstein equation for the potential have
a very complex multi-periodic structure, which can be used for comparison with observational data.

 \section{Conclusions}

  The emergence of two-dimensional structures such as Zeldovich pancakes is associated with the density perturbations described by
  classical or (weakly) relativistic hydrodynamics. 
        Recent observational tensions, such as the Hubble constant tension, can require the consideration of nonidentical processes that lead to structure formation and evolution on local and global cosmological scales.
        
        Thus, we have considered a description of the local Universe, taking into account (i) the self-consistent many-particle interaction by means of the Vlasov kinetic formalism and (ii) modified gravity with the cosmological constant term from Eq. (1), that is to say, the repulsion at galaxy cluster scales. This description predicts two Hubble flows, local and global ones, with nonidentical Hubble parameters. Thus, in a certain sense, the Hubble tension can be considered as empirical support of Eq. (1) and of the presented mechanism of structure formation.  
        
Based on these aspects, we have developed a kinetic model for the occurrence of voids separated by two-dimensional surfaces, following from a
rigorous analysis of the Poisson equation and its quasi-oscillatory solutions. We show the appearance mechanism of voids of different scales, depending on the local region of the Universe.

Its appearance in the Poisson equation with the modified potential following from the theorem in \citet{G1} on a general function for
the "sphere-point" identity poses a mathematical problem of the study of inhomogeneous Fredholm integral equations.

The principal consequence of the developed mechanism is that the cosmological constant poses a natural scaling for the voids, via the characteristic distance when the second, repulsive $\Lambda$ term dominates over the first term in Eq. (1): 
\begin{equation}
r_{cr} = (\frac{3 GM}{\Lambda c^2})^{1/3}. 
\end{equation}This distance, $r_{cr}$, together with the total mass within the corresponding volume, or, equivalently, with the mean density in that volume, will determine the actual size of the voids \citep{GS7}.

We can illustrate this quantitatively with the available observational parameters of the Virgo Supercluster and the Laniakea Supercluster. For the Virgo Supercluster, one has M = $1.48 \times 10^{15} M_{\odot}$ \citep{Ein,Reid}, and the scale at which the repulsive $\Lambda$ term can lead to a local Hubble flow is in the range $17.3<r< 18.4\, Mpc$. For the Laniakea Supercluster, one has the mass $M = 10^{17} M_\odot$ \citep{Tul} and the critical radius when the repulsion overwhelms the gravitational attraction, $r_{cr} \simeq 51 Mpc$. These parameters imply diameters of voids from $35\, Mpc$ up to $100\, Mpc$ in such environments. 

These predictions can be directly tested in observational surveys, namely, via the search for correlations in the sizes of the voids versus the total masses and mean densities of their local regions  (e.g., \citealt{Ce,spot,S1,S2}). Such studies, along with the dynamics and the flows of galaxy clusters and superclusters, can be instrumental in revealing whether there are additional tensions in descriptions of the late and early Universe.  

\section{Acknowledgments}

We are thankful to  the referee for helpful comments.

\end{document}